\documentclass[10pt,a4paper]{article}
\usepackage[utf8]{inputenc}
\usepackage[english]{babel}
\usepackage{amsmath}
\usepackage{amsfonts}
\usepackage{amssymb}
\usepackage{amsthm}
\usepackage{color}
\usepackage{float}
\usepackage{dsfont}

\usepackage{tikz}	%include TikZ-engine
\usetikzlibrary{calc}

\usepackage[left=2cm,right=2cm,top=2cm,bottom=2cm]{geometry}

\numberwithin{equation}{section}

\newtheorem{lem}{Lemma}
\newtheorem{thm}{Theorem}
\newtheorem{cor}{Corollary}

\theoremstyle{remark}

\newcommand{\R}{\mathbb{R}}
\newcommand{\Sp}{\mathbb{S}}
\newcommand{\eqdef}{\overset{def}{=}}	%eqdef sign

\newcommand{\tP}{\mathcal{P}}
\newcommand{\supp}{\mathrm{supp}}
\newcommand{\sgn}{\mathrm{sgn}}

\title{An example of non-uniqueness for Radon transforms\\ 
	with continuous positive rotation invariant weights}
\author{F.O. Goncharov\thanks{CMAP, Ecole Polytechnique, CNRS, Universit\'{e} Paris-Saclay, 91128, Palaiseau, France; \newline \indent email: fedor.goncharov.ol@gmail.com} 
	\and	 
		R. G. Novikov$^*$\footnote{IEPT RAS, 117997  
			Moscow, Russia;
			\newline\indent email: roman.novikov@polytechnique.edu}}
\begin{document}
\maketitle
\abstract{We consider weighted Radon transforms $R_W$
along hyperplanes in $\R^3$ with strictly positive weights $W$.  We construct an example of such a transform with non-trivial kernel $\mathrm{Ker}R_W$ in the space of infinitely smooth compactly supported functions and with continuous weight. Moreover, in this example the weight $W$ is rotation invariant. In particular, by this result we continue studies of Quinto (1983), Markoe, Quinto (1985), Boman (1993) and Goncharov, Novikov (2017). We also extend our example to the case of weighted Radon transforms along two-dimensional planes in $\R^d, \, d\geq 3$.\\

\noindent \textbf{Keywords:} Radon transforms, integral geometry, injectivity, non-injectivity\\

\noindent \textbf{AMS Mathematics Subject Classification:} 44A12, 53C65, 65R32
}

\section{Introduction}
%Weighted Radon transforms are of interest for their intrinsic properties \cite{guillemin1976radon}, \cite{boman1987support}, \cite{gel2014integral}, \cite{helgason1990totally} and applications  
%to tomography \cite{cormack1963representation}, \cite{tretiak1980exponential}, 
%\cite{novikov2002inversion}, \cite{deans2007radon}.  
\par We consider weighted Radon transforms $R_W$ in $\R^d$ defined by 
\begin{align}\label{rad.def}
	&R_Wf(s,\theta) =  \int\limits_{x\theta=s}W(x,\theta)f(x)\, dx,\\ \nonumber
	&s\in \R, \, \theta\in\Sp^{d-1}, \, x\in \R^d, \, d\geq 2, 
\end{align}
where $W=W(x,\theta)$ is the weight, $f = f(x)$ is a test function on $\R^d$.
\par We assume that 
\begin{align}\label{w.assump}
	W = \overline{W} \geq c > 0, \, W\in L^{\infty}(\R^d\times\Sp^{d-1}),
\end{align}
where $\overline{W}$ denotes the complex conjugate of $W$, $c$ is a constant.
\par The aforementioned transforms $R_W$ arise in various domains of pure and applied mathematics; see, e.g.,
\cite{beylkin1984inversion}, \cite{beylkin1985imaging}, 
\cite{boman1987support}, \cite{boman1993example}, \cite{davidfinch1986}, \cite{lavrientiev1973first}, 
\cite{goncharov2016analog}, \cite{goncharov2016iterative}, \cite{goncharov2017example}, 
\cite{kunyansky1992generalized}, \cite{natterer2001mathematics},  
\cite{novikov2014weighted},  \cite{quinto1983invertibility}, \cite{quinto1983invertibilityerrat} and references therein.
\par In particular, studies on the transforms $R_W$ under assumptions \eqref{w.assump} were recently 
continued in \cite{goncharov2016analog}, \cite{goncharov2016iterative}, \cite{goncharov2017example} for $d\geq 3$.
\par Note that the works \cite{goncharov2016analog}, \cite{goncharov2016iterative} extend to the case of $R_W$,  $d\geq 3$, the two-dimensional injectivity and reconstruction results of \cite{kunyansky1992generalized}, \cite{novikov2011weighted}, \cite{novikov2014weighted},  \cite{guillement2014inversion}.
\par On the other hand, under assumptions \eqref{w.assump}, the work \cite{goncharov2017example} gives 
an example of $R_W$, $d\geq 3$, with non-trivial kernel in $C_0^{\infty}(\R^d)$ (infinitely smooth functions with compact support). This example was constructed in \cite{goncharov2017example} proceeding from 
the example of non-uniqueness of \cite{boman1993example} for $R_W$ in $\R^2$ and a recent result of \cite{goncharov2016analog}.

\par In the two-dimensional example of non-uniqueness of \cite{boman1993example} the weight $W$ satisfies 
\eqref{w.assump}, for $d=2$, and is infinitely smooth everywhere. In the multidimensional example of 
non-uniqueness of \cite{goncharov2017example} the weight satisfies \eqref{w.assump}, for $d\geq 3$, is 
infinitely smooth almost everywhere but is not yet continuous at some points.
\par In the present work we construct an example of $R_W$, for $d=3$, with non-trivial kernel in $C_0^{\infty}(\R^3)$, where $W$ satisfies \eqref{w.assump} and is continuous everywhere. Moreover, in this example $W$ is rotation invariant and $R_Wf \equiv 0$ for some non-zero spherically symmetric $f\in C_0^{\infty}(\R^3)$.
\par The rotation invariancy  of the latter example is its principal advantage in comparison with the aforementioned examples of \cite{boman1993example} and \cite{goncharov2017example}.
\par By our rotation invariant example of non-uniqueness we also continue studies of \cite{markoe1985elementary}, where an example of non-uniqueness for $R_W$ was constructed for $d=2$. In the example of \cite{markoe1985elementary} the weight $W$ is bounded and positive but is not yet continuous and strictly positive. The continuity and strict positivity of $W$ is the principal advantage of the example of the present work in comparison with the example of \cite{markoe1985elementary}.
%\par In the present work we construct an example of $R_W$ in $\R^3$, under assumptions of \eqref{w.assump}, with non-trivial kernel $\mathrm{Ker}\, R_W$ in $C_0^{\infty}(\R^3)$ (infinitely smooth functions with compact support).
%\par Moreover, in this example $W$ is continuous on $\R^3\times \Sp^2$, rotation invariant (see formulas \eqref{w.rot.inv}, \eqref{u.sym.def} below) and $R_Wf \equiv 0$ for some non-zero spherically symmetric $f\in C_0^{\infty}(\R^3)$.
%\par By this result we continue studies of \cite{markoe1985elementary}, where
%related example with bounded $W$ was constructed in 2D. We emphasize that the weight $W$
%is continuous in our 3D-example in contrast with the 2D-example of \cite{markoe1985elementary}.
\par Following \cite{quinto1983invertibility} we say that $W$ is rotation invariant if 
and only if
\begin{equation}\label{w.rot.inv}
 W(x,\theta) = U(|x-(x\theta)\theta|, x\theta), \, x\in \R^d, \, \theta\in \Sp^{d-1},
\end{equation}
for some positive and continuous $U$ such that 
\begin{align} \label{u.sym.def}
	&U(r,s) = U(-r,s) = U(r,-s), \, r\in \R, \, s\in \R.
\end{align}
\par On the other hand, we recall that weighted Radon transforms $R_W$ in $\R^d$ with smooth weights $W$ satisfying properties \eqref{w.assump}, \eqref{w.rot.inv},  \eqref{u.sym.def} are injective for $f\in L_0^2(\R^d)$ (square integrable functions on $\R^d$ with compact support); see \cite{quinto1983invertibility}. Here the smoothness of $W$ can be specified, at least, as $C^1$ for $d=2$ and $d=3$. 
In view of the aforementioned counterexamples of \cite{markoe1985elementary} and of the present work, some smoothness of $W$ is crucial for these injectivity results. 
\par In the present work we also extend our rotation invariant example of non-uniqueness for $R_W$ in $\R^3$ to 
the case of weighted Radon transforms $R^{d,2}_{W}$ along two-dimensional planes in $\R^d, \, d>3$; see Section~\ref{real_extension}. In this case $R^{d,2}_Wf$ is defined on $\tP^{d,2}$ (manifold of all oriented two-dimensional planes in $\R^d$) and is overdetermined already. That is 
\begin{equation}
\dim \tP^{d,2} = 3d-6 > \dim \R^d = d \text{ for } d>3. 
\end{equation}
Nevertheless, $R^{d,2}_Wf\equiv 0$ on $\tP^{d,2}$ in our result.
\par We expect that the results of the present work admit generalizations to the weighted Radon transforms 
$R_{W}^{d,n}$ along $n$-dimensional planes in $\R^d$ for arbitrary $d$ and $n$ such that $1 \leq n < d, \, d\geq 2$. For $n=1$ such results are already obtained in \cite{gonnov2017example}.
\par Note also that the construction of the present work was developed in a large extent in the process of adopting the Boman's construction of the aforementioned work \cite{boman1993example}.
\par In Section~\ref{sect.prelim} we give some preliminaries.
\par Our main results are formulated in detail in Sections~\ref{sect.main},~\ref{real_extension}.
\par Proofs are given in Sections~\ref{prv.lm1}-\ref{pr.lm5}.

\section{Some preliminaries}\label{sect.prelim}
\paragraph{Notations for $d=3$.} Let 
\begin{align}
&B = \{x\in \R^3 : |x| < 1\}, \overline{B} = \{x\in \R^3 : |x|\leq 1\},\\
&\mathcal{P} = \R\times \Sp^2,\\ \label{pr.e0}
&\mathcal{P}_0(\delta) = \{(s,\theta) \in \mathcal{P} : |s| > \delta \},\\ \label{pr.e1}
&\mathcal{P}_1(\delta) = \mathcal{P}\backslash \mathcal{P}_0(\delta) = \{(s,\theta) \in \mathcal{P} : 
|s| \leq \delta\}, \, \delta > 0,\\ \label{pr.lambda}
&\mathcal{P}(\Lambda) = \{(s,\theta)\in \mathcal{P} : s\in \Lambda\}, \, 
\Lambda \subset \R, \\ \label{loc.weight.domain}
&\Omega(\Lambda) = \{(x,\theta)\in \R^3\times \Sp^2 : x\theta\in \Lambda\}, \,\Lambda
\subset \R,\\
\label{int.def}
&\mathcal{J}_{s,\varepsilon} = \mathcal{J}_{|s|,\varepsilon} = 
(-|s|-\varepsilon,-|s|+\varepsilon) \cup (|s|-\varepsilon, |s|+\varepsilon)
\subset \R, \, s\in \R, \, \varepsilon > 0.
\end{align}
\par Note that $\mathcal{P}_0(\delta), \mathcal{P}_1(\delta)$ of \eqref{pr.e0}, \eqref{pr.e1} are particular cases of $P(\Lambda)$ of formula \eqref{pr.lambda}. 
\par In addition, we interpret $\mathcal{P}$ as the set of all oriented planes in $\R^3$. If $P = (s,\theta)\in \mathcal{P}$, then 
\begin{equation}
	P = P_{(s,\theta)} = \{x\in \R^3 : x\theta =s \} \text{ (modulo orientation)}
\end{equation}
and $\theta$ gives the orientation of $P$ (in the sense that ordered tuple $(e_1, e_2, \theta)$ is positively oriented in $\R^3$ with any orthonormal positively oriented basis $e_1,e_2$ on $P$).
\par The set $\mathcal{P}_0(\delta)$ in \eqref{pr.e0} is considered as the set of all oriented planes in $\R^3$ which are positioned at distance greater than $\delta$ from the origin.
\par The set $\mathcal{P}_1(\delta)$ in \eqref{pr.e1} is considered as the set of all oriented planes in $\R^3$ which are located at distance less or equal than $\delta$.

\paragraph{Rotation invariancy for $d=3$.} 
\par Symmetries \eqref{w.rot.inv}, \eqref{u.sym.def} of $W$ can be also written as
\begin{align}\label{w.new.sym}
	&W(x,\theta) = \tilde{U}(|x|, x\theta), \, x\in \R^3, \, \theta\in \Sp^2,\\ 
	&\tilde{U}(r,s) = \tilde{U}(r, -s), 
	\, \tilde{U}(r,s) = \tilde{U}(-r,s),\, r\in\R, \, s\in \R, \label{u.new.sym}
\end{align}
where $\tilde{U}$ is positive and continuous on $\R\times \R$. 
Using the formula $|x|^2 = |x\theta|^2 + |x-(x\theta)\theta|^2, \, \theta\in \Sp^2$, one can see that symmetries \eqref{w.rot.inv}, \eqref{u.sym.def} and symmetries \eqref{w.new.sym}, \eqref{u.new.sym} of $W$ are equivalent.

\paragraph{Additional notations.}
\par For a function $f$ on $\R^d$ we denote its restriction to a subset $\Sigma$ by $f|_\Sigma$.
\par By $C_0, \, C_0^{\infty}$ we denote continuous compactly supported and infinitely smooth compactly supported functions, respectively.

\paragraph{Partition of unity.}
We recall the following classical result (see Theorem 5.6 in \cite{do1992riemannian}):\\ 
{\itshape
	Let $\mathcal{M}$ be a $C^{\infty}$-manifold, which is Hausdorff and satisfies second countability axiom (i.e. has countable base). Let also $\{U_i\}_{i=1}^{\infty}$ be the open locally-finite cover of $\mathcal{M}$.
	\par Then there exists a $C^{\infty}$-smooth locally-finite partition of unity $\{\psi_i\}_{i=1}^{\infty}$ on $\mathcal{M}$,
	such that 
	\begin{equation}\label{gen.part.prop}
		\supp\, \psi_i\subset U_i.
	\end{equation}	 
}
\par In particular, any open interval $(a,b)\subset \R$ and $\mathcal{P}\simeq \R\times \Sp^2$ satisfy conditions of the aforementioned statement. It will be used in Subsection~\ref{s.w1.constr}.

\section{Main results for $d=3$}\label{sect.main}
\begin{thm}\label{main.thm}
	There exist a non-zero spherically symmetric function $f\in C_0^\infty(\R^3)$ with support in $\overline{B}$, and $W$ satisfying \eqref{w.assump}-\eqref{u.sym.def} such that 
	\begin{equation}\label{m.thm.stm.}
		R_Wf\equiv 0, 
	\end{equation}
	where $R_W$ is defined in \eqref{rad.def}.
\end{thm}
\par The construction of $f$ and $W$ proving Theorem~\ref{main.thm} is presented below in this section. This construction adopts considerations of \cite{boman1993example}. 
In particular, we construct $f$, first, and then $W$. 

\subsection{Construction of $f$}
The function $f$ is constructed as follows:
\begin{align}\label{m.def.ser.}
	&f = \sum\limits_{k=1}^{\infty}\dfrac{f_k}{k!},\\ \label{m.def.fk}
	&f_k(x) = f_k(|x|) = \Phi(2^k(1-|x|))\cos(8^k|x|^2), x\in \R^3, \, k=1,2, \dots,
\end{align}
for arbitrary $\Phi\in C^{\infty}(\R)$ such that
\begin{align}\label{phi.suppint}
	&\supp\, \Phi = [4/5, 6/5], \\
	\label{phi.supp}
	&0 < \Phi(t) \leq 1 \text{ for } t\in (4/5,6/5), \\
	\label{phi.unity.cond}
	&\Phi(t) = 1 \text{ for } t\in [9/10, 11/10].
\end{align}

\par Properties \eqref{phi.suppint}, \eqref{phi.supp} imply that functions $f_k$ in \eqref{m.def.fk} have disjoint supports and series \eqref{m.def.ser.} converges for every fixed $x\in \R^3$. 
\begin{lem}\label{b.lem.ch.sgn}
	Let $f$ be defined by \eqref{m.def.ser.}-\eqref{phi.unity.cond}. Then $f$ is spherically symmetric, $f\in C_0^{\infty}(\R^3)$ and $\supp \, f \subseteq \overline{B}$. In addition, if $P\in \mathcal{P}, \, P\cap B\neq \emptyset$, then $f|_P\not\equiv 0$ and $f|_P$ has non-constant sign.
\end{lem}
\par Lemma~\ref{b.lem.ch.sgn} is proved in Section~\ref{prv.lm1}.

\subsection{Construction of $W$}
\par In our example $W$ is of the following form:
\begin{align}\label{b.W.form}
\begin{split}
	W(x,\theta) &= \sum\limits_{i=0}^{N}\xi_i(|x\theta|)W_i(x,\theta) \\
	& = 
	\xi_0 (|x\theta|) W_0(x,\theta) + \sum\limits_{i=1}^{N}\xi_i(|x\theta|)
	W_i(x,\theta),\, x\in \R^3, \, \theta\in
	\Sp^2, 
\end{split}
\end{align}
where
\begin{align}
\label{b.part.unit}
\{\xi_i(s&), \, s\in \R\}_{i=0}^{N} \text{ is a $C^{\infty}$-smooth partition of unity on }\R,\\
\label{b.un.part.prop}
\xi_i(s) &= \xi_i(-s), \, s\in \R, \, i=\overline{0,N},
\end{align}
\vspace{-0.5cm}
\begin{align}
\label{b.weight.desc}
\begin{split}
&W_i(x,\theta) \, \text{are bounded continuous strictly positive and }\\
&\text{rotation invariant (according to 
\eqref{w.rot.inv}, \eqref{u.sym.def}) on } \supp\, \xi_i(|x\theta|), \, i = \, \overline{0,N}, \text{ respectively}.
\end{split}
\end{align}
\par From \eqref{b.W.form}-\eqref{b.weight.desc} it follows that $W$ of \eqref{b.W.form} satisfies the conditions \eqref{w.assump}-\eqref{u.sym.def}.

\par The weight $W_0$ is constructed in Subsection~\ref{s.w0.constr} and has the following properties:
\begin{align}\label{w0.c.rinv}
	&W_0\text{ is bounded, continuous and rotation invariant on } \{(x,\theta) : |x\theta| > 1/2\},\\
	\label{w0.pos.d}
	\begin{split}
		&\text{there exists $\delta_0\in (1/2,1)$ such that: } \\
		&\qquad \qquad W_0(x,\theta) \geq 1/2 \text{ if } |x\theta| > \delta_0,\\
		&\qquad \qquad W_0(x,\theta) = 1 \quad \text{ if } |x\theta|\geq 1,
	\end{split}\\ \label{w0.zero}
	&R_{W_0}f(s,\theta) = 0 \text{ for } |s| > 1/2, \, \theta \in \Sp^2,
\end{align}
where $R_{W_0}$ is defined according to \eqref{rad.def} for $W = W_0$, $f$ is given by 
\eqref{m.def.ser.}, \eqref{m.def.fk}.
\par In addition, 
\begin{align}\label{supp.xi0}
	&\supp \, \xi_0 \subset (-\infty, -\delta_0)\cup (\delta_0, +\infty),\\
	\label{super.prop.xi0}
	&\xi_0(s) = 1 \text{ for } |s|\geq 1,
\end{align}
where $\delta_0$ is the number of \eqref{w0.pos.d}.
\par In particular, from \eqref{b.part.unit}, \eqref{w0.pos.d}, \eqref{supp.xi0} it follows that
\begin{equation}\label{w0.positiv}
	W_0(x,\theta)\xi_0(|x\theta|) > 0 \text{ if } \xi_0(|x\theta|) > 0.
\end{equation}
\vspace{-0.8cm}
\paragraph{Remark 1.} The result of \eqref{w0.c.rinv}-\eqref{w0.zero} can be considered as a counterexample to the Cormack-Helgason support theorem (see Theorem 3.1 in \cite{natterer2001mathematics}) in the framework of the theory of weighted Radon transforms
under assumptions \eqref{w.assump} and even under assumptions \eqref{w.assump}-\eqref{u.sym.def}.\\
\par In addition,
\begin{align}\label{w.loc.prop}
	&\xi_i(|x\theta|) W_i(x,\theta) \text{ are bounded, continuous and rotation invariant on }
	\R^3\times \Sp^2,\\ 
	&W_i(x,\theta)\geq 1/2 \text{ if }  \xi_i(|x\theta|) \neq 0,\\
	&R_{W_i}f(s,\theta) = 0 \text{ if } \xi_i(|s|)\neq 0,\\ \label{w.loc.prop.last}
	&i = \overline{1,N}, \, x\in \R^3, \, \theta\in \Sp^2, \, s\in \R.
\end{align}
Weights $W_1, \dots, W_N$ of \eqref{b.W.form} and $\{\xi_i\}_{i=0}^{N}$ are constructed in Subsection~\ref{s.w1.constr}.
\par Result of Theorem~\ref{main.thm} follows from Lemma~\ref{b.lem.ch.sgn} and formulas \eqref{b.W.form}-\eqref{b.un.part.prop}, \eqref{w0.c.rinv}-\eqref{w0.zero}, \eqref{w0.positiv}-\eqref{w.loc.prop.last}. 
\par We point out that the construction of $W_0$ of \eqref{b.W.form} is substantially different from the construction of $W_1, \dots , W_N$. 
In particular, the weight $W_0$ is defined on the planes $P\in \mathcal{P}$ which can be close to the boundary $\partial B$ of $B$ which results in  restrictions on the smoothness of $W_0$.

\subsection{Construction of $W_0$}\label{s.w0.constr}
Let 
\begin{equation}\label{psik.unit.part.def}
	\{\psi_k\}_{k=1}^{\infty} \text{ be a } C^{\infty} \text{ partition of unity on 
	$(1/2, 1)$, such that }
	\supp\, \psi_k\subset (1-2^{-k+1}, 1-2^{-k-1}), \, k\in \mathbb{N}.
\end{equation}
\par Note that
\begin{equation}
	1-2^{-(k-2)-1} < 1 -2^{-k}(6/5), \, k\geq 3.
\end{equation}
Therefore,
\begin{equation}\label{supp.coincide}
	\forall s_0,\,t_0\in \R : s_0\in \supp\, \psi_{k-2}, \, t_0\in \supp\, \Phi(2^k(1-t))
	\Rightarrow s_0 < t_0, \, k\geq 3.
\end{equation}
\par The weight $W_0$ is defined  by the following formulas
\begin{align}\label{b.w0.def}
	&W_0(x,\theta) = 
	\begin{cases}
		1- G(x,\theta)\sum\limits_{k=3}^{\infty}k!f_k(x)
		\dfrac{\psi_{k-2}(|x\theta|)}{H_k(x,\theta)}, \, 1/2 < |x\theta| < 1,\\
		1, \, |x\theta| \geq 1
	\end{cases}, \\ \label{b.gkhk.def}
	&G(x,\theta) = \int\limits_{y\theta=x\theta}
	\hspace*{-0.2cm}f(y)\, dy, \, 
	H_k(x,\theta) = \int\limits_{y\theta=x\theta}
	\hspace*{-0.2cm}f_k^2(y)\, dy,\\ \nonumber
	&x\in \R^3,\, \theta\in \Sp^2,
\end{align}
where $f_k$ are defined in \eqref{m.def.fk}.
\par Formula \eqref{b.w0.def} implies that $W_0$ is defined on $\mathcal{P}_0(1/2)\subset \mathcal{P}$. Due to \eqref{m.def.fk} and \eqref{supp.coincide}, in \eqref{b.gkhk.def} we have that $H_k(x,\theta) \neq 0$ if $\psi_{k-2}(|x\theta|)\neq 0$.
\par Also, for any fixed $(x,\theta)\in \R^3\times \Sp^2, \, 1/2<|x\theta|<1$, the series in the right hand-side of \eqref{b.w0.def} has only a finite number of non-zero terms (in fact, no more than two) and, hence, $W_0$ is well-defined.
\par By the spherical symmetry of $f$, functions $G, H_k$ in \eqref{b.w0.def} are of the type \eqref{w.new.sym}, \eqref{u.new.sym}. Therefore, $W_0$ is rotation invariant (in the sense \eqref{w.new.sym}, \eqref{u.new.sym}). 
\par Actually, formula \eqref{w0.zero} follows from \eqref{m.def.ser.}-\eqref{phi.suppint}, \eqref{b.w0.def}, \eqref{b.gkhk.def} (see Subsection~\ref{subs.w0.zero} for details).
\par Using the construction of $W_0$ and the assumption that $|x\theta| > 1/2$ (implying 
that $\mathrm{sign}(x\theta)$ is locally constant) one can see that $W_0$ is $C^{\infty}$ 
on its domain of definition, possibly, except points with $|x\theta|=1$.

\begin{lem}\label{b.cont.lem}
	Let $W_0$ be defined by \eqref{b.w0.def}, \eqref{b.gkhk.def}. Then the following estimate holds:
	\begin{align}\label{w.w0.conv}
		&|1-W_0(x,\theta)| \leq C_0\rho(|x\theta|)
		\left(\log_2 \dfrac{1}{\rho(|x\theta|)}\right)^4,  \\ 
		&W_0(x,\theta) \rightarrow 1 \text{ as }|x\theta|\rightarrow 1,\\ \nonumber
		&x\in \R^3, \, \theta\in \Sp^2, \, 1/2 < |x\theta| < 1. 
	\end{align}	
	where $\rho = \rho(s) = 1 - s, \, s\in (1/2,1)$, $C_0$ is a positive constant depending on $\Phi$.
\end{lem}
\par Lemma~\ref{b.cont.lem} is proved in Section~\ref{prv.lm2}.
\par The result of Lemma~\ref{b.cont.lem} completes the proof of \eqref{w0.pos.d}.
\par This completes the description of $W_0$ and $\delta_0$.

\subsection{Construction of $W_1, \dots, W_N$ and $\xi_0,\dots, \xi_N$}\label{s.w1.constr}
\begin{lem}\label{b.lem.loc}
	Let $f\in C_0(\R^3)$ be spherically symmetric, $P_{(s_0,\theta_0)}\in \mathcal{P}$, 
	$f|_{P_{(s_0,\theta_0)}}\not\equiv 0$ and 
	$f|_{P_{(s_0,\theta_0)}}$ changes the sign. Then:
	\begin{enumerate} 
		\item[(i)] there exist $\varepsilon > 0$ and weight $W_{f, s_0, \varepsilon}$
	such that
	\begin{align}\label{b.loc.weight}
		R_{W_{f,s_0,\varepsilon}}f(s,\theta) = 0 
		 \text{ for } s\in \mathcal{J}_{s_0,\varepsilon}, \, 
   		 \theta\in \Sp^2, 
	\end{align}
	where $\mathcal{J}_{s,\varepsilon}$ is defined in \eqref{int.def}, 
	 $W_{f,s_0,\varepsilon}$ is defined on the open set 
	$\Omega(J_{s_0,\varepsilon})$, defined by \eqref{loc.weight.domain};
	\item[(ii)] weight $W_{f,s_0,\varepsilon}$ is bounded, continuous, strictly positive
		and rotation invariant on  $\Omega (\mathcal{J}_{s_0,\varepsilon})$.
	\end{enumerate}
\end{lem}
\par Lemma~\ref{b.lem.loc} is proved in Section~\ref{prv.lm3}.
\par Let $f$ be the function of \eqref{m.def.ser.}, \eqref{m.def.fk}. Then, using Lemmas~\ref{b.lem.ch.sgn},~\ref{b.lem.loc} one can see that 
\begin{align}\label{exist.loc.weight}
	\begin{split}
	&\forall \delta\in (0,1) \text{ there exist } 
	\{J_i = \mathcal{J}_{s_i,\varepsilon_i}, W_i= W_{f, s_i, \varepsilon_i}\}_{i=1}^N\\
	&\text{such that } J_i, \, i=\overline{1,N}, \text{ is an open cover of } 
	[-\delta,\delta]\\
	&\text{and } W_i \text{ satisfy (i) and (ii) (of Lemma~\ref{b.lem.loc}) on } \Omega(J_i). 
	\end{split}
\end{align}
\par Actually, we consider  \eqref{exist.loc.weight} for the case of $\delta = \delta_0$ of \eqref{w0.pos.d}.
\par Note that in this case $\{\mathcal{P}(J_i)\}_{i=1}^N$ for $J_i$ of \eqref{exist.loc.weight} is an open cover of $\mathcal{P}_1(\delta_0)$. 

\par  To the set $\mathcal{P}_0(\delta_0)$ we associate the open set
\begin{equation}\label{choice.v0}
J_0= (-\infty, \delta_0) \cup (\delta_0, +\infty)\subset \R.
\end{equation}
Therefore, the collection of intervals $\{J_i, \, 
i = \overline{0,N}\}$ is an open cover of $\R$. 
\par We construct the partition of unity $\{\xi_i\}_{i=0}^N$ on $\R$ as follows: 
\begin{align}\label{constr.supp.def}
	&\xi_i(s) = \xi_i(|s|) = \dfrac{1}{2}(\tilde{\xi}_i(s) + \tilde{\xi}_i(-s)), \, 
	s\in \R,\\ \label{constr.supp.prop}
	&\supp \, \xi_i \subset J_i, \, i = \overline{0,N},
\end{align}
where $\{\tilde{\xi}_i\}_{i=0}^{N}$ is a partition of unity for the open cover $\{J_i\}_{i=0}^N$ (see Section~\ref{sect.prelim}, Partition of unity, for $U_i = J_i$).

\par Properties \eqref{supp.xi0}, \eqref{constr.supp.prop} follow from \eqref{gen.part.prop} for $\{\tilde{\xi}_i\}_{i=0}^N$ (with $U_i = J_i$), the symmetry of $J_i = \mathcal{J}_{s_i,\varepsilon_i}, \, 
i = \overline{1,N}$, choice of $J_0$ in \eqref{choice.v0} and from \eqref{constr.supp.def}.
 
In addition, \eqref{super.prop.xi0} follows from \eqref{choice.v0} and the construction of $J_i, \, 
i = \overline{1,N}$, from \eqref{exist.loc.weight} (see the proof of Lemma~\ref{b.lem.loc} and properties \eqref{exist.loc.weight} in Section~\ref{prv.lm3} for details). 
\par Properties \eqref{w.loc.prop}-\eqref{w.loc.prop.last} follow from  \eqref{exist.loc.weight} for $\delta = \delta_0$ and from \eqref{choice.v0}-\eqref{constr.supp.prop}.

\section{Extension to the case of $R_W^{d,2}$}\label{real_extension}
We consider the weighted Radon transforms $R_{W}^{d,2}$ along two-dimensional planes in $\R^d$, 
defined by 
\begin{align}
	\begin{split}\label{wRt_two_dim_planes}
	R_{W}^{d,2}f(P) = \int\limits_{P}W(x,P)f(x)\, dx, \, P\in \tP^{d,2}, \, x\in P, \, d\geq 3,
	\end{split}
\end{align} 
where $W = W(x,P)$ is the weight, $f=f(x)$ is a test function on $\R^d$,
\begin{align}
	\tP^{d,2}\text{ is the manifold of all oriented two-dimensional planes } P \text{ in } \R^d.
\end{align}
\par Note that the transform $R_W^{d,2}$ is reduced to $R_W$ of \eqref{rad.def} for $d=3$.
\par We say that $W$ in \eqref{wRt_two_dim_planes} is rotation invariant if and only if 
\begin{align}
	&W(x,P) = \tilde{U}(|x|, \mathrm{dist}(P,\{0\})),\\
	&\tilde{U}(r,s) = \tilde{U}(r, -s), 
	\, \tilde{U}(r,s) = \tilde{U}(-r,s),\, r\in\R, \, s\in \R,
\end{align}
where $\tilde{U}$ is some positive and continuous function on $\R\times \R$, $\mathrm{dist}(P,\{0\})$ denotes the distance from the origin $\{0\}\in \R^d$ to the plane $P$. Note that $W(x,P)$ is independent of the orientation of $P$ in this case.
\par Consider $\tilde{U}$ and $\tilde{f}$ such that
\begin{align}\label{tilde_func_def}
	W(x,\theta) = \tilde{U}(|x|, |x\theta|), \, f(x) = \tilde{f}(|x|), \, x\in \R^3, \, \theta\in \Sp^2,
\end{align}
for $W$ and $f$ of Theorem~\ref{main.thm} of Section~\ref{sect.main}.
\par Theorem~\ref{main.thm} implies the following corollary:
\begin{cor}\label{extend_corollary}
	Let $W$ and $f$ be defined as
	\begin{align}\label{wext_def}
		&W(x,P) = \tilde{U}(|x|, \mathrm{dist}(P, \{0\})), \, P\in \tP^{d,2}, \, x\in P, \\ \label{fext_def}
		&f(x) = \tilde{f}(|x|), \, x\in \R^d,
	\end{align}
	where $\tilde{U}, \, \tilde{f}$ are the functions of \eqref{tilde_func_def}, $d > 3$. Then
	\begin{equation}\label{extended_zero}
		R_W^{d,2}f\equiv 0 \text{ on } \tP^{d,2}.
	\end{equation}
	In addition, the weight $W$ is continuous strictly positive and rotation invariant, $f$ is 
	infinitely smooth compactly supported on $\R^d$ and $f\not\equiv 0$. 
\end{cor}
Formula \eqref{extended_zero} is proved as follows:
\begin{align}\label{extend_integral}
	&R_W^{d,2}(P) = \int\limits_{P}\tilde{U}(|x|, \mathrm{dist}(P, \{0\}))\tilde{f}(|x|)\, dx = 
				 I \eqdef \int\limits_{P'}\tilde{U}(|x|,s)\tilde{f}(|x|)\,dx,\\  \nonumber
				 &P' = \{s e_3 + x_1e_1 + x_2e_2: x = (x_1,x_2) \in \R^2\}, \, 
				 s = \mathrm{dist}(P, \{0\}), 
\end{align}
where $(e_1,\dots,e_d)$ is the standard basis in $\R^d$. In addition, $I = 0$ by Theorem~\ref{main.thm}. 
\par Properties of $W$ and $f$ mentioned in Corollary~\ref{extend_corollary} follow from definitions \eqref{wext_def}, \eqref{fext_def} and properties of $\tilde{U}$ and $\tilde{f}$ (arising in Theorem~\ref{main.thm}).

\section{Proofs of Lemma~\ref{b.lem.ch.sgn} and formula \eqref{w0.zero}}\label{prv.lm1}
\subsection{Proof of Lemma~\ref{b.lem.ch.sgn}}
The spherical symmetry of $f$ follows from \eqref{m.def.ser.}, \eqref{m.def.fk}.\\
The series in \eqref{m.def.ser.} converges uniformly with all  derivatives of $f_k$. Therefore, $f\in C^{\infty}(\R^3)$. Due to \eqref{m.def.ser.},
\eqref{m.def.fk}, \eqref{phi.suppint}, \eqref{phi.supp} we have that $\supp \, f_k\subset B, \, k\geq 1$. Therefore, $\supp\, f\subseteq \overline{B}$.
\par It remains to show that $f$ restricted to any straight line $l$ in $\R^3$ intersecting $B$ changes the sign. This implies change of the sign for $f|_P$ for any plane $P$ such that 
$P\cap B \neq \emptyset$.
\par We consider 
\begin{align}\label{supp.fk}
	&D_k = \{x\in \R^3 : |x|\in (1-2^{-k}(6/5), 1 -2^{-k}(4/5))\}, \, k \geq 1,\\ 
	\label{pr.lin.def}
	&l(x_0,\omega) = \{x\in \R^3 : x = x(t) = x_0 + \omega t,\,  t\in \R\}, \omega\in \Sp^2,\, x_0\in \R^3, \, x_0\omega = 0.
\end{align}
\par Note that $\supp\, f_k = \overline{D}_k \subset B$. Note also that the line $l(x_0,\omega)$ intersects $B$ if and only if $|x_0| < 1$. 
\par Assuming that
\begin{equation}\label{pr.cond.lin}
	|x_0| < 1-2^{-k}(6/5),
\end{equation}
we consider $D_k\cap l(x_0,\omega) = I^{-}_k\sqcup I^{+}_k$ (see Figure 1):
\begin{align}
	&I^{-}_k = \{x(t) : t\in (-t_1, -t_0)\},\\
	&I^{+}_k = \{x(t) : t\in (t_0, t_1)\}, \\ \nonumber
	&t_0 : = t_0(k), \, t_1 : = t_1(k).
\end{align}
\begin{figure}[H]
\centering
\begin{tikzpicture}[scale=0.7]
	%draw circles
	\draw [thick,domain=40:140] plot ({5*cos(\x)}, {5*sin(\x)});
	\draw [thick,domain=40:140] plot ({3.5*cos(\x)}, {3.5*sin(\x)});
	\draw [dotted] (0,0) -- (40:5cm);
	\draw [dotted] (0,0) -- (140:5cm);
	%draw coords
	\draw [dotted] (0,-1) -- (0, 5.5);
	\draw [dotted] (-4,0) -- (4,0);
	%draw the main line
	\draw [->] (0,3.5) -- ++(30:4cm);
	\draw  (0,3.5) -- ++(210:3cm);
	\draw [dotted, ->] (0,0) -- (120:3cm);
	\draw [->] (0,0) -- (75:4.3cm);
	\draw [thick,domain=30:90] plot ({0 + 0.3*cos(\x)}, {3.5+ 0.3*sin(\x)});
	
	%\draw text
	\node [right] at (3.5,5.5) {$l(x_0,\omega)$};
	\node [right] at (-1.3,1.7) {$x_0$};
	\node at (0.2, 4.0) {$\gamma$};
	\node at (-0.35, 3.8) {$t_0$};
	\node at (1.9, 5.0) {$t_1$};
	\node at (1.1,2.0) {$x(t)$};
	\node at (-3.3, 2.8) {$\mathrm{supp}\, f_k$};
	\node at (1.7, 4.0) {$I_k^+$};
	\node at (4.2, 2.2) {$1-2^{-k}(6/5)$};
	\node at (5.3, 3.2) {$1-2^{-k}(4/5)$};
	\node [right] at (0,0) {$\scriptstyle 0$}; 
	\node at (1, -1.5) {Figure 1.};
\end{tikzpicture}
\end{figure}
\par One can see that assumption \eqref{pr.cond.lin} holds for all $k\geq k_0(|x_0|) = -\ln
 \left(\frac{5}{6}(1-|x_0|)\right)$.
\par By the Cosine theorem we have (see Figure 1):
\begin{equation}\label{pr.varph.def}
	\varphi(t) : = |x(t)|^2 = (t-t_0)^2 + |x(t_0)|^2 - 2|x(t_0)|(t-t_0)\cos(\pi - \gamma) 
	\text{ for } t\in [t_0,t_1].
\end{equation} 
One can see also that 
\begin{equation}\label{cos_neg}
	\gamma\in [0,\pi/2], \, \cos(\pi -\gamma) \leq 0.
\end{equation}
\par Let 
\begin{equation}\label{pr.gk.def}
	g_k(t) := \cos(8^k \varphi(t)), \, t\in [t_1,t_2],
\end{equation}
where $\varphi(t)$ is defined in \eqref{pr.varph.def}.
\par It is sufficient to show that $g_k$ changes the sign on $(t_0,t_1)$ for sufficiently large $k$. 
\par Due to \eqref{m.def.ser.}-\eqref{phi.supp} this implies that $f$ changes the sign on $I^{+}_k$.
\par From \eqref{pr.varph.def}, \eqref{cos_neg} we obtain the inequality
\begin{equation}
	\varphi'(t) = 2(t-t_0) - 2|x(t_0)|\cos(\pi-\gamma) \geq 0 \, 
	\text { for } t\in (t_0,t_1),\, \gamma\in [0,\pi/2],
\end{equation}
which implies the phase in \eqref{pr.gk.def} is monotonously increasing on $t\in (t_0, t_1)$.
\par The full variation $V_{(t_0,t_1)}(\varphi)$ of the monotonous phase $\varphi(t)$ on $ (t_0,t_1)$ is given by the formula
\begin{equation}\label{pr.phase.var}
	V_{(t_0,t_1)}(\varphi) = (t_1-t_0)^2 - 2|x(t_0)|(t_1-t_0)\cos(\pi -\gamma).
\end{equation}
\par From \eqref{cos_neg}, \eqref{pr.phase.var} we obtain the following inequality
\begin{align}\label{pr.full.var}
	V_{(t_0,t_1)}(\varphi) \geq (t_1-t_0)^2.
\end{align}
From \eqref{pr.gk.def}, \eqref{pr.full.var} it follows that $g_k$ changes the sign on $t\in (t_0, t_1)$, for example, if 
\begin{equation}\label{pr.cond.change}
	8^k V_{(t_0,t_1)}(\varphi) \geq 2\pi \text{ or } (t_1-t_0) \geq \sqrt{2\pi}4^{-k}.
\end{equation}
\par On the other hand, $(t_1-t_0)$ is exactly the length of the segment $I^{+}_k$ (see Figure 1). 
Therefore, 
\begin{equation}\label{pr.supp.time}
(t_1-t_0) \geq (2/5)2^{-k}.
\end{equation}
\par Inequality \eqref{pr.supp.time} implies that \eqref{pr.cond.change} holds for $k \geq 3$. Therefore, $g_k$ of \eqref{pr.gk.def} changes the sign on $(t_0,t_1)$ starting from $k \geq \max (3, k_0(|x_0|))$.
\par Lemma~\ref{b.lem.ch.sgn} is proved.

\subsection{Proof of formula \eqref{w0.zero}}\label{subs.w0.zero}
\par From \eqref{rad.def}, \eqref{m.def.ser.}-\eqref{phi.supp}, \eqref{psik.unit.part.def}, \eqref{b.w0.def}, \eqref{b.gkhk.def} it follows that:
\begin{align}
	R_{W_0}f(s,\theta) &= \int\limits_{x\theta=s}
	\hspace*{-0.2cm}f(x)\, dx - 
	G(s\theta,\theta)
	\sum\limits_{k=3}^{\infty}k!\psi_{k-2}(|s|)
	\dfrac{\int\limits_{x\theta=s}f(x)f_k(x)dx}
	{H_k(s\theta,\theta)} \\
	&= \int\limits_{x\theta=s}\hspace*{-0.2cm}f(x)\, dx - 
	\int\limits_{x\theta=s}\hspace*{-0.2cm}f(x)dx
	\sum\limits_{k=3}^{\infty}\psi_{k-2}(|s|)
	\dfrac{\int\limits_{x\theta=s}f_k^2(x)dx}
	{\int\limits_{x\theta=s}f_k^2(x)\,dx}\\
	&= \int\limits_{x\theta=s}\hspace*{-0.2cm}f(x)\, dx - 
	\int\limits_{x\theta=s}\hspace*{-0.2cm}f(x)dx
	\sum\limits_{k=3}^{\infty}\psi_{k-2}(|s|) = 0, \, |s| > 1/2, \, \theta\in \Sp^2.
\end{align}
\par Formula \eqref{w0.zero} is proved.

\section{Proof of Lemma~\ref{b.cont.lem} }\label{prv.lm2}
\par Let
\begin{equation}\label{plane.assump}
	\Lambda_k := \{(x,\theta)\in \R^3\times \Sp^2: |x\theta|\in(1-2^{-k+3}, 1-2^{-k+1})\}, k\in \mathbb{N}, \, k\geq 4.
\end{equation}
From \eqref{psik.unit.part.def} it follows that, for $k\geq 4$:
\begin{align}\label{supp.psi.1}
	&\supp \, \psi_{k-1} \subset  (1-2^{-k+2}, 1-2^{-k}),\\ \label{supp.psi.2}
	&\supp \, \psi_{k-2} \subset  (1-2^{-k+3}, 1-2^{-k+1}), \\ \label{supp.psi.3}
	&\supp \, \psi_{k-3} \subset  (1-2^{-k+4}, 1-2^{-k+2}). 
\end{align}
Formulas \eqref{b.w0.def}, \eqref{b.gkhk.def},
 \eqref{supp.psi.1}-\eqref{supp.psi.3} imply the following expression for $W_0(x,\theta)$:
	\begin{equation}\label{main_rho0_estim}
	\begin{split}
		W_0(x,\theta) = 1 - G(x,\theta)
		&\left(
			(k-1)!f_{k-1}(x)\dfrac{\psi_{k-3}(|x\theta|)}{H_{k-1}(x,\theta)}  \right. \\ 
		&\left. 
		 	+k!f_{k}(x)\dfrac{\psi_{k-2}(|x\theta|)}{H_{k}(x,\theta)} \right. \\
		&\left. 
			+(k+1)!f_{k+1}(x)\dfrac{\psi_{k-1}(|x\theta|)}{H_{k+1}(x,\theta)}
		\right), \, (x,\theta)\in \Lambda_k, \, k\geq 4.
	\end{split}
	\end{equation}

\begin{lem}\label{main_lemm}
There are positive constants $c_1,c_2, k_1$ depending on $\Phi$, such that
\begin{align}\label{estimate_func}
	&\,|f_k(x)| \leq c_1, \text{ for } k\in \mathbb{N},\\ \label{est_H_k_term}
	&\left|\dfrac{\psi_{k-2}(|x\theta|)}{H_k(x,\theta)}\right| \leq 
	c_2 2^{k} \quad  \text{  for  } k\geq k_1 \text{ and } |x\theta|\leq 1-2^{-k+1},\\
	\label{estimate_G}
	&\,|G(x,\theta)| \leq c_1 \dfrac{4^{-k}}{k!}\quad \,\,\,\, \text{ for } 
	k\geq 3 \text{ and } |x\theta|\geq 1-2^{-k},\\ 
	\nonumber
	&\,x\in \R^3, \, \theta\in \Sp^2, 1/2 < |x\theta| < 1,
\end{align}
where $f_k, G, H_k$ are defined in \eqref{m.def.fk}, \eqref{b.gkhk.def}.
\end{lem}
\par Lemma~\ref{main_lemm} is proved in Section~\ref{pr.lm5}.

	From definition \eqref{plane.assump} and estimates \eqref{est_H_k_term}, \eqref{estimate_G} it 	follows that 
	\begin{align}\label{inequality_G}
		&|G(x,\theta)| \leq c_1 4^{-k+3}/(k-3)!, \\
		\label{inequality_Hk}
		&\left|\dfrac{\psi_{k-2}(|x\theta|)}{H_k(x,\theta)}\right| \leq 
	c_2 2^{k},\\ \nonumber
		&\text{ for }(x,\theta) \in \Lambda_k,\, k\geq \max(4, k_1).
	\end{align}
	In addition, properties \eqref{supp.psi.1}-\eqref{supp.psi.3} and estimate \eqref{est_H_k_term} 
	imply that:	
	\begin{align}
	  \label{ineq_left}
	  &\begin{cases}
		\psi_{k-1}(|x\theta|) = 0,\\
		\left|
			\dfrac{\psi_{k-3}(|x\theta|)}{H_{k-1}(x,\theta)}
		\right| \leq c_22^{k-1}
	   \end{cases} \text{ if } \quad |x\theta|\in (1-2^{-k+3}, 1-2^{-k+2}),\\
	   &\begin{cases}\label{ineq_right}
	   		\psi_{k-2}(|x\theta|)=0,\\
	   		\left|
			\dfrac{\psi_{k-1}(|x\theta|)}{H_{k+1}(x,\theta)}
			\right| \leq c_22^{k+1}
	   \end{cases}
	   \text{ if } \quad |x\theta|\in (1-2^{-k+2}, 1-2^{-k+1}),\\
	   &\begin{cases}\label{ineq_mid}
	   		\psi_{k-1}(|x\theta|) = 0,\\
	   		\psi_{k-3}(|x\theta|)=0\\
	   \end{cases}
	   \quad \quad \quad \, 
	   \text{ if } \quad |x\theta|=1-2^{-k+2},
	\end{align}
	for $(x,\theta)\in \Lambda_k, \, k\geq \max(4,k_1)$.
	\par Note that the condition $(x,\theta)\in \Lambda_k$ is splitted into 
	the assumptions of \eqref{ineq_left}, \eqref{ineq_right}, \eqref{ineq_mid}.
	\par Due to formulas \eqref{main_rho0_estim}, \eqref{inequality_G}-\eqref{ineq_mid},
	we obtain the following estimates:
	\begin{align}\label{big.estim.left}\begin{split}
		|1-W_0(x,\theta)| &= |G(x,\theta)|
		\left|
			(k-1)!f_{k-1}(x)\dfrac{\psi_{k-3}(|x\theta|)}{H_{k-1}(x,\theta)} + 
			k!f_{k-2}(x)\dfrac{\psi_{k-2}(|x\theta|)}{H_{k}(x,\theta)}
		\right|\\
		&\leq c_14^{-k+3}(c_1c_2(k-2)(k-1) 2^{k-1} + c_1c_2(k-2)(k-1)k 2^{k})\\
		&\leq 2^6c_1^2c_22^{-k}k^3 \quad \text{ if } \quad
		|x\theta|\in (1-2^{-k+3}, 1-2^{-k+2}),
	 	\end{split}
	\end{align}
	\begin{align}\label{big.estim.right}
		\begin{split}
		|1-W_0(x,\theta)| &= |G(x,\theta)|
		\left|
			k!f_{k}(x)\dfrac{\psi_{k-2}(|x\theta|)}{H_{k}(x,\theta)}
			+(k+1)!f_{k+1}(x)\dfrac{\psi_{k-1}(|x\theta|)}
			{H_{k+1}(x,\theta)}
		\right|\\
		&\leq c_14^{-k+3}(c_1c_22^k(k-1)(k-2) + c_1c_22^{k+1}(k-2)(k-1)k(k+1))\\
		&\leq 2^{12}c_1^2c_22^{-k}k^4 \quad \text{ if } \quad
		|x\theta|\in (1-2^{-k+2}, 1-2^{-k+1}),
		\end{split}
	\end{align}
	\begin{align}\label{big.estim.mid}
	\begin{split}
		|1-W_0(x,\theta)| &= |G(x,\theta)|
		\left|
			k!f_{k}(x)\dfrac{\psi_{k-2}(|x\theta|)}{H_{k}(x,\theta)}
		\right|\hspace{4.2cm}\\
		&\leq 2^6c_1^2c_22^{-k}k^3	\quad \text{ if } \quad
		|x\theta|=1-2^{-k+2}.
	\end{split}
	\end{align}
	Estimates \eqref{big.estim.left}-\eqref{big.estim.mid} imply that 
	\begin{align}\label{discr.estim}
		|1-W_0(x,\theta)|\leq C\cdot  2^{-k}k^4, \, (x,\theta)\in \Lambda_k, \, 
		k\geq \max(4,k_1).
	\end{align}
	where $C$ is a positive constant depending on $c_1,c_2$ of
	Lemma~\ref{main_lemm}. 
	\par In addition, for $(x,\theta)\in \Lambda_k$ we have that $2^{-k+1}< \rho(|x\theta|) < 2^{-k+3}$, which together with \eqref{discr.estim} imply \eqref{w.w0.conv}.
	\par Lemma~\ref{b.cont.lem} is proved.

\section{Proof of Lemma~\ref{b.lem.loc}}\label{prv.lm3}
\par Let $(e_1,e_2)$ be an orthonormal basis on $P_{(s,\theta)}\in \mathcal{P}$ and the origin of the coordinate system on $P_{(s,\theta)}$ is located 
at $s\theta\in P_{(s,\theta)}$.
\par By $u = (u_1,u_2), \, u\in \R^2$, we denote the coordinates on $P_{(s,\theta)}$ with respect to $(e_1,e_2)$.
\par Using Lemma~\ref{b.lem.ch.sgn} one can see that
\begin{equation}\label{plane.symm}
	f|_{P_{(s,\theta)}}\in C_0^{\infty}(\R^2), \, f|_{P_{(s,\theta)}}(u) = f|_{P_{(s,\theta)}}(|u|), \, u\in \R^2.
\end{equation}
\par By our assumptions $f|_{P_{(s_0,\theta_0)}}(u)$ changes the sign. 
\par Using this assumption and \eqref{plane.symm} 
one can see that there exist $\psi_{1,s_0}, \psi_{2,s_0}$, such that:
\begin{align}\label{psi.loc.symm}
	&\psi_{1,s_0}\in C([0,+\infty)), \, \psi_{1,s_0} \geq 0, 
	\psi_{2,s_0}(u) := \psi_{1,s_0}(|u|), \, 
u \in \R^2,
	\\ \label{non_zero_psi_int}
	&\int\limits_{P_{(s_0,\theta_0)}} 
	\hspace{-0.3cm} f \psi_{2,s_0}\, d\sigma \neq 0.
\end{align}
and if 
\begin{equation} \label{loc.weight.nonvan}
	\int\limits_{P_{(s_0,\theta_0)}}\hspace*{-0.3cm}f\,d\sigma \neq 0
\end{equation}
then also 
\begin{align}
	\label{sgn.choice}
	&\sgn
	(
	\hspace*{-0.3cm}
		\int\limits_{P_{(s_0,\theta_0)}} \hspace*{-0.35cm} f\, d\sigma
	)\, 
	\sgn
	(	\hspace*{-0.3cm}
		\int\limits_{P_{(s_0,\theta_0)}} \hspace*{-0.35cm} f\psi_{2,s_0}\, d\sigma 
	) = -1,
\end{align}
where $d\sigma = du_1\, du_2$ (i.e., $\sigma$ is the standard Euclidean measure on $P_{(s,\theta)}$).
\par Let 
\begin{equation}\label{loc.weight}
	W_{f, s_0}(x,\theta) = 1-\psi_{1,s_0}(|x-(x\theta)\theta|)
	\dfrac{\int\limits_{P_{(x\theta,\theta)}}\hspace{-0.3cm}
	f\, d\sigma}{\int_{P_{(x\theta,\theta)}}\hspace{-0.1cm}f\psi_{2,s_0}\, d\sigma},
	\, x\in \R^3, \, \theta\in \Sp^2,
\end{equation}
where $d\sigma = du_1\,du_2$ and $(u_1,u_2)$ are the 
coordinates on $P_{(s,\theta)}, \, s = x\theta$, defined at the beginning of this proof.
\par Results of Lemma~\ref{b.lem.ch.sgn} and property \eqref{psi.loc.symm} imply that
\begin{equation}\label{save.sign}
	\int\limits_{P_{(x\theta,\theta)}}\hspace{-0.3cm}
	f\, d\sigma \text{ and }\hspace{-0.2cm}
	\int\limits_{P_{(x\theta,\theta)}}\hspace{-0.3cm}
	f\psi_{2,s_0}\, d\sigma \text{ depend only on }
	|x\theta|, \text{ where } x\in \R^3, \, \theta\in \Sp^2.
\end{equation}
\par From \eqref{loc.weight}, \eqref{save.sign} it follows that $W_{f, s_0}$ is rotation-invariant in the sense \eqref{w.rot.inv}, \eqref{u.sym.def}.
\par Formulas \eqref{non_zero_psi_int}, \eqref{loc.weight}, \eqref{save.sign} and   
properties of $f$ and $\psi_{2,s_0}$ of Lemma~\ref{b.lem.ch.sgn} and \eqref{psi.loc.symm} imply that
\begin{align}\label{ws0.contin0}
	\exists \varepsilon_1 > 0 : \int\limits_{P_{(x\theta,\theta)}}\hspace*{-0.3cm}f\psi_{2,s_0}\,d\sigma \neq 0, \text{ for }  
	(x,\theta)\in \Omega(\mathcal{J}_{s_0,\varepsilon_1}),
\end{align} 
where the sets $\mathcal{J}_{s,\varepsilon}, \Omega(\mathcal{J})$ are defined in 
\eqref{loc.weight.domain}, \eqref{int.def}, respectively.
\par In addition, using \eqref{loc.weight}, \eqref{ws0.contin0}, one can see that
\begin{align}\label{ws0.contin}
	W_{f, s_0} \text{ is continuous on } (x,\theta)\in \Omega(\mathcal{J}_{s_0,\varepsilon_1}).
\end{align}
\par In addition, from \eqref{plane.symm}-\eqref{save.sign} it follows that 
\begin{align}\nonumber
	\text{ if } |x\theta| = |s_0| \text{ then } W_{f, s_0}(x,\theta) &= 1-
	\psi_{1,s_0}(|x-(x\theta)\theta|)
	\dfrac{\int\limits_{P_{(s_0,\theta)}}\hspace{-0.3cm}
	f\, d\sigma}{\int_{P_{(s_0,\theta)}}f\psi_{2,s_0}\, d\sigma} \\ \label{origin.pos}
	&= 1-\psi_{1,s_0}(|x-(x\theta)\theta|)
	\dfrac{\int\limits_{P_{(s_0,\theta_0)}}\hspace{-0.3cm}
	f\, d\sigma}{\int_{P_{(s_0,\theta_0)}}f\psi_{2,s_0}\, d\sigma} \geq 1.
\end{align}

From properties of $f,\psi_{1,s_0}, \psi_{2,s_0}$ of Lemma~\ref{b.lem.ch.sgn} and 
 \eqref{psi.loc.symm} 
and from formulas \eqref{loc.weight}, \eqref{save.sign}, \eqref{ws0.contin}, \eqref{origin.pos} it follows that 

\begin{equation}\label{pos.cond.zero}
	\exists \varepsilon_0 > 0\, (\varepsilon_0 < \varepsilon_1) : W_{f, s_0}(x,\theta) \geq 1/2, \text{ for }
	(x,\theta)\in \Omega(\mathcal{J}_{s_0,\varepsilon_0}),
\end{equation}
which implies strict positiveness for $W_{f, s_0}$ on $\Omega(
\mathcal{J}_{s_0,\varepsilon})$.
\par Properties \eqref{save.sign}, \eqref{ws0.contin}, \eqref{pos.cond.zero} imply 
item (ii) of Lemma~\ref{b.lem.loc} for $W_{f,s_0,\varepsilon} : = W_{f,s_0}$, defined on 
$\Omega(\mathcal{J}_{s_0,\varepsilon_0})$.
\par From \eqref{rad.def}, \eqref{loc.weight}, \eqref{ws0.contin0} it follows that
\begin{align} \nonumber
	R_{W_{f,s_0}}f(s,\theta) &= \int\limits_{P_{(s,\theta)}}\hspace{-0.2cm}
	W_{f,s_0}(\cdot, \theta)f\, d\sigma
	\\ \label{loc_weight_zero_ident}
	&=\int\limits_{P_{(s,\theta)}}\hspace{-0.2cm}f\, d\sigma \, - 
	\dfrac{\int\limits_{P_{(s,\theta)}}\hspace{-0.3cm}
	f\, d\sigma}{\int_{P_{(s,\theta)}}f\psi_{2,s_0}\, d\sigma}	
	\int\limits_{P_{(s,\theta)}}\hspace{-0.3cm} f \psi_{2,s_0}\, d\sigma = 0
	\, \text{ for } s\in \mathcal{J}_{s_0,\varepsilon_0}, \theta\in \Sp^2.
\end{align}
Item (i) of Lemma~\ref{b.lem.loc} follows from \eqref{loc_weight_zero_ident}.
\par Lemma~\ref{b.lem.loc} is proved.

\section{Proof of Lemma~\ref{main_lemm}}\label{pr.lm5}
\subsection{Proof of estimate \eqref{estimate_func}}
\par Estimate \eqref{estimate_func} follows from \eqref{m.def.fk} and 
properties \eqref{phi.suppint}-\eqref{phi.unity.cond}.
\subsection{Proof of estimate \eqref{estimate_G}}
\par From definitions \eqref{m.def.ser.}, \eqref{b.gkhk.def} we have that
\begin{align}\label{G_def}
	&G = \sum\limits_{k=1}^{\infty} \dfrac{G_k}{k!},\\
	&\label{Gk_def}
	G_k(x,\theta) = \int\limits_{y\theta = x\theta}\hspace*{-0.2cm} f_k(y)\,  dy, \, x\in \R^3, \, 
	\theta\in \Sp^2.	
\end{align}
\par Parametrization of the points $y(r,\phi)$ on $P_{(s,\theta)}\in \mathcal{P}, 
\, s\in \R, \, \theta\in \Sp^2$, is given by the formula 
\begin{align}\label{y_param}
	&y(r,\phi) = s\theta + r(e_1\cos \phi + e_2\sin \phi), \,	r\in [0, + \infty), \, \phi\in [0,2\pi], 
\end{align}
where $(e_1,e_2)$ is some fixed orthonormal basis on  $P_{(s,\theta)}$.
\par On the other hand, 
\begin{equation}\label{r_gamma_param}
	r = r(\gamma) = |s|\tan (\gamma), \, \gamma\in [0,\pi/2), 
\end{equation}
where $\gamma$ is the angle between $s\theta$ and the radius-vector $y(r, \phi)$ of \eqref{y_param}.
\par It is convenient to rewrite $y(r,\phi)$ of \eqref{y_param} as $y = 
y(r(\gamma),\phi) \eqdef y(\gamma, \phi), \, \gamma\in [0,\pi/2), \, \phi\in [0,2\pi]$.
\par The standard Lebesgue measure $\sigma$ on $P_{(s,\theta)}$ is given by the following formula: 
\begin{align}\nonumber
	d\sigma (\gamma,\phi) = r(\gamma,\phi) d\phi \, dr(\gamma) &= |s|\tan \gamma\,  d\phi \, dr(\gamma)\\ \label{plain_mes}
	 &= |s|^2 \dfrac{\sin\gamma}{\cos^3\gamma}\, d\phi\, d\gamma.
\end{align}
From \eqref{m.def.fk}, \eqref{Gk_def}-\eqref{plain_mes} we obtain
\begin{align}\nonumber
	G_k(x,\theta) &= s^2\int\limits_{0}^{2\pi}d\phi\int\limits_{0}^{\pi/2} 
	\Phi\left(2^{k}\left(1-\frac{|s|}{\cos\gamma}\right)\right) \cos \left(
		8^k \dfrac{|s|^2}{\cos^2\gamma}
	\right)\dfrac{\sin\gamma}{\cos^3\gamma}\, d\gamma\\ \nonumber
	&=-2\pi|s|^2\int\limits_{0}^{\pi/2} 
	\Phi\left(2^{k}\left(1-\frac{|s|}{\cos\gamma}\right)\right) \cos \left(
		8^k \dfrac{|s|^2}{\cos^2\gamma}
	\right)\dfrac{d(\cos\gamma)}{\cos^3\gamma}\\ \nonumber
	&=\{t=\cos\gamma\}=-2\pi|s|^2 \int\limits_{1}^{0}
	\Phi\left(2^{k}\left(1-\frac{|s|}{t}\right)\right) \cos \left(
		8^k \dfrac{|s|^2}{t^2}
	\right)\dfrac{dt}{t^3}\\ \label{G_k_value}
	&= \{u=\frac{1}{t^2}\} = \pi|s|^2\int\limits_{1}^{+\infty} 
	\Phi(2^k(1-|s|\sqrt{u})) \cos(8^k|s|^2u) \, du, \, s= x\theta.
\end{align}
\par From \eqref{phi.suppint}--\eqref{phi.unity.cond}, \eqref{G_k_value} it follows that
\begin{align}\nonumber
	G_{k}(x,\theta) &= 8^{-k}\pi\int\limits_{1}^{+\infty}
	\Phi (2^k(1-|s|\sqrt{u}))
	\,d\left(
		\sin (8^k|s|^2u)
	\right)\\  \label{G_k_1prtintgr}
	&= 8^{-k}\pi\left(-
		\Phi(2^{k}(1-|s|))\sin(8^k|s|^2) - 
		\int\limits_{1}^{+\infty}\left(\dfrac{d}{du} \Phi(2^k(1-|s|\sqrt{u}))\right) 
		\sin(8^k |s|^2u)\, du
	\right),\\ \label{phi.unif.bound}
	&|\Phi(2^k(1-|s|))\sin(8^k|s|^2)| \leq 1, \\
	\nonumber
	&\left|
		\int\limits_{1}^{+\infty}\left(\dfrac{d}{du} \Phi(2^k(1-|s|\sqrt{u}))\right) 
		\sin(8^k |s|^2u)\, du
	\right|
	\leq 2^k\max_{t\in \R}|\Phi'(t)|
	\int\limits_{\Lambda_{k,|s|}} \hspace{-0.3cm} du\\ \label{unif.prt.int.bound}
	&\qquad \qquad \qquad \qquad \qquad \qquad \qquad 
	\qquad \qquad \quad \, \, \, \leq 
	2^k\max_{t\in \R}|\Phi'(t)|,\\
	&\Lambda_{k,|s|} = \{u\geq 1 :  2^k(1-|s|\sqrt{u}) \in [4/5, 6/5] \},\\
	\nonumber
	&\text{where } 1/2 < |s| < 1, \, s= x\theta, \,  k\in \mathbb{N}.
\end{align}
\par Note that 
\begin{equation}\label{labmda_length}
	|\Lambda_{k,|s|}| \leq 1 \text{ for } 1/2 < |s| < 1,
\end{equation}
where $|\Lambda|$ denotes the length of $\Lambda$.
\par Formulas \eqref{G_k_1prtintgr}-\eqref{labmda_length} imply that
\begin{equation}\label{G_k_estimate}
	|G_k(x,\theta)| \leq 4^{-k}\pi \max_{t\in\R}|\Phi'(t)| \text{ for }
	 1/2 < |s| < 1, \, s = x\theta, \,  k\in \mathbb{N}.
\end{equation}
\par Note that for $y\in P_{(s,\theta)}$, the following inequality holds:\begin{equation}\label{cross.estim}
	2^k(1 - |y|) \leq 2^k(1-|s|) \leq 2^{k-m} \leq 4/5
	\text{ for } 1-2^{-m} \leq |s|  < 1, \, k < m , \, m\geq 3.
\end{equation}
\par Formulas \eqref{m.def.fk}, \eqref{phi.suppint}, \eqref{cross.estim} imply that
\begin{equation}\label{G_k_prop}
	P_{(s,\theta)}\cap \supp\, f_k = \emptyset \text{ if } 
	|s| \geq 1-2^{-m}, \, k < m.
\end{equation}
In turn, \eqref{Gk_def}, \eqref{G_k_prop} imply that 
\begin{equation}\label{G_k_prop2}
	G_k(x,\theta) = 0 \text{ for } k < m, \, |x\theta| \geq 1-2^{-m}.
\end{equation}
\par Due to \eqref{G_def}, \eqref{G_k_estimate}, \eqref{G_k_prop2} we have that:
\begin{align}\nonumber
	\begin{split}
	|G(x,\theta)| &\leq \sum\limits_{k=1}^{\infty} |G_k(x,\theta)|/k! = \sum\limits_{k=m}^{\infty} |G_k(x,\theta)|/k! \\
	&\leq \max_{t\in\R}|\Phi'(t)|\pi4^{-m} / m! \sum\limits_{k=0}^{\infty} 4^{-k}
	= c_1 \dfrac{4^{-m}}{m!}, \, c_1 = \dfrac{4\pi}{3}\max_{t\in \R}|\Phi'(t)|
	\end{split}\\ \label{G_k_final_estim}
	&\text{for } |x\theta| \geq 1-2^{-m}, \, m\geq 3.
\end{align} 
\par Estimate \eqref{estimate_G} follows from \eqref{G_k_final_estim}.

\subsection{Proof of estimate \eqref{est_H_k_term}} 
\par  For each $\psi_k$ from \eqref{psik.unit.part.def} we have that:
\begin{equation}
	|\psi_k| \leq 1.
\end{equation}
Therefore, it is sufficient to show that 
\begin{equation}
	H_k \geq C_22^{-k} \text{ for } k\geq k_1, \, C_2 = c_2^{-1}.
\end{equation}
\par Due to formula \eqref{b.gkhk.def} and in a similar way with \eqref{G_k_value} we obtain 
\begin{align} \label{Hk_def_integr}
	H_k(x,\theta) &= |s|^2\pi\int\limits_{1}^{\infty}\Phi^2(2^k(1-|s|\sqrt{u})) \cos^2(8^k|s|^2u)\, du = H_{k,1}(x,\theta) + H_{k,2}(x,\theta), \, 
	s= x\theta,
	\\ 
	\label{Hk1_def}
	H_{k,1}(x,\theta) &= \dfrac{\pi|s|^2}{2}\int\limits_{1}^{+\infty} \Phi^2(2^k(1-|s|\sqrt{u}))\, du, \\
	\label{Hk2_def} 
	H_{k,2}(x,\theta) &= \dfrac{\pi|s|^2}{2}\int\limits_{1}^{+\infty}
	\Phi^2(2^k(1-|s|\sqrt{u})) \cos(2\cdot 8^k|s|^2u) \, du.
\end{align} 
Note that 
\begin{align}\label{secnd.assump}
	&2^k(1-|s|) \geq 2^{k}\cdot 2^{-k+1} \geq 2 > 6/5 \text{ for }
	|s|\leq 1-2^{-k+1}, \, k \geq 3.
\end{align}
In turn, \eqref{phi.suppint}, \eqref{secnd.assump} 
	imply that
\begin{align} \label{Hk_supp.condition}
	&\Phi(2^k(1-|s|\sqrt{u})) = 0 \text{ for } u \leq 1, \, 
	|s|\leq 1-2^{-k+1}, \, k \geq 3.
\end{align}
\par Using \eqref{secnd.assump} one can see that 
\begin{align}\label{u1u2_exist}
	&\exists u_1\geq 1, u_2\geq 1, \,  u_2 > u_1  \text{ such that } \begin{cases}
		2^k(1-|s|\sqrt{u_1}) = 11/10,\\
		2^k(1-|s|\sqrt{u_2}) = 9/10,
	 \end{cases}\\
	\label{u1u2_length}
	&|u_2-u_1| \geq (\sqrt{u_2} - \sqrt{u_1}) = \frac{2^{-k}}{5}
	|s|^{-1} \geq \frac{2^{-k}}{5}, \\ \nonumber
	&\text{for } 1/2 < |s|\leq 1-2^{-k+1}, \, k \geq 3.
\end{align}
\par Using \eqref{phi.suppint}, \eqref{phi.unity.cond}, \eqref{Hk1_def}, \eqref{Hk_supp.condition}, \eqref{u1u2_length} we obtain 
\begin{align}\label{mean_val_thm}
	&H_{k,1}(x,\theta) \geq  \dfrac{\pi}{8}
	\int\limits_{u_1}^{u_2}du \geq 2^{-k}\dfrac{\pi}{40}, \, \, 
	\text{ for }\, \, 1/2 < |x\theta| < 1-2^{-k+1}, \, k\geq 3.
\end{align}

\par On the other hand, using \eqref{phi.suppint}, \eqref{Hk2_def}, \eqref{Hk_supp.condition}, in a similar way with  \eqref{G_k_1prtintgr}-\eqref{labmda_length},  
we obtain
\begin{align}\label{last_bound}
	 \begin{split}
	 |H_{k,2}(x,\theta)| = 
	\dfrac{\pi|s|^2}{2}
	\left|\int\limits_{1}^{+\infty}
	\right.&
	\left.
	\Phi^2
	(2^k(1-|s|\sqrt{u})) 
	\cos(2\cdot 8^k|s|^2u) \, du
	\right| \\
	& =  \dfrac{\pi}{4}8^{-k}|s|^{-2}
	\left|
		\int\limits_{1}^{+\infty} \sin (2\cdot 8^{k}|s|^2u)
		\left(\dfrac{d}{du}\Phi^2(2^k(1-|s|\sqrt{u}))\right) du 
	\right|\\
	&\leq \dfrac{\pi}{4}8^{-k}|s|^{-1}
	\max_{t\in \R}|\Phi(t)|\cdot \max_{t\in \R}|\Phi'(t)|
	\cdot 2^k\int\limits_{\Lambda_{k,|s|}} du\\
	&\leq \dfrac{\pi}{2}4^{-k} \max_{t\in \R}|\Phi(t)|\cdot \max_{t\in \R}|\Phi'(t)|,\,  s = x\theta,
	\end{split}\\ \nonumber
	\text{for } 1/2 < |x\theta| < 1-&2^{-k+1}, \, k\geq 3. 
\end{align}
From \eqref{Hk_def_integr}-\eqref{Hk2_def}, \eqref{mean_val_thm}, \eqref{last_bound} it follows that 
\begin{align}\label{H_ineq}
\begin{split}
	|H_k(x,\theta)| &\geq |H_{k,1}(x,\theta)| - |H_{k,2}(x,\theta)| \\ 
	&\geq \dfrac{\pi}{40} 2^{-k} - \dfrac{\pi}{2}4^{-k}
	\max_{t\in \R}|\Phi(t)|\cdot \max_{t\in \R}|\Phi'(t)|\\
	&\geq C_2 2^{-k} \, \text{ for } 1/2 < |x\theta| < 1-2^{-k+1}, \, k\geq 
	k_1,\\
	C_2 &= \dfrac{\pi}{40} - 2^{-k_1}\dfrac{\pi}{2} 
		\max_{t\in \R}|\Phi(t)|
		\max_{t\in \R}|\Phi'(t)|,
\end{split}
\end{align}
where $k_1$ is arbitrary constant such that $k_1 \geq 3$ and $C_2$ is positive. 
\par Estimate \eqref{est_H_k_term} follows from \eqref{H_ineq}.
\par Lemma~\ref{main_lemm} is proved.

\section*{Aknowledgements}
This work is partially supported by the PRC $n^{\circ}$ 1545 CNRS/RFBR: \'{E}quations quasi-lin\'{e}aires, probl\`{e}mes inverses et leurs applications. 

%\bibliography{boman_3D.bib}

\begin{thebibliography}{}

\end{thebibliography}


\begin{thebibliography}{Bom85}


\bibitem[Bey84]{beylkin1984inversion}
Beylkin, G., 
\newblock{The inversion problem and applications of the generalized Radon transform}.
\newblock{ \em Communications on pure and applied mathematics}, 
37(5) : 579-599, 1984.

\bibitem[Bey85]{beylkin1985imaging}
Beylkin, G.,
\newblock{Imaging of discontinuities in the inverse scattering problem by inversion of a causal generalized Radon transform}
\newblock{ \em Journal of Mathematical Physics}, 26(1) : 99-108, 1985.

\bibitem[BQ87]{boman1987support}
Boman, J., Quinto, E.T.,
\newblock { Support theorems for real-analytic {R}adon transforms}.
\newblock {\em Duke Mathematical J.}, 55(4):943-948, 1987.


\bibitem[Bom93]{boman1993example}
Boman, J.,
\newblock { An example of non-uniqueness for a generalized Radon transform}.
\newblock {\em Journal d’Analyse Mathematique}, 61(1):395--401, 1993.

\bibitem[MCar92]{do1992riemannian}
Do Carmo, M. P., 
\newblock{Riemannian Geometry}.
\newblock Birkhäuser Basel, 1992.


\bibitem[Fi86]{davidfinch1986}
Finch, D.,
\newblock{Uniqueness for the attenuated X-ray transform in the physical range}.
\newblock{\em Inverse problems}, 2(2), 1986.

\bibitem[GN16]{goncharov2016analog}
Goncharov, F.O.,  Novikov, R.G., 
\newblock An analog of {C}hang inversion formula for weighted {R}adon
  transforms in multidimensions.
\newblock {\em Eurasian Journal of Mathematical and Computer Applications}, 4(2):23-32, 2016.

\bibitem[Gon17]{goncharov2016iterative}
Goncharov, F.O., An iterative inversion of weighted Radon transforms along hyperplanes.
\newblock {\em arXiv:1611.10209v8}, 2017.

\bibitem[GN17]{goncharov2017example}
Goncharov, F.O.,  Novikov, R.G.,
\newblock An example of non-uniqueness for the weighted Radon transforms along
hyperplanes in multidimensions.
\newblock {\em arXiv:1709.04194v2}, 2017.

\bibitem[GonNov17]{gonnov2017example}
Goncharov, F.O., Novikov, R.G., 
\newblock A breakdown of injectivity for weighted ray transforms in multidimensions. 
hal-01635188, version 1, 2017.

\bibitem[GuiNov14]{guillement2014inversion}
Guillement, J.-P.,  Novikov, R.G.,
\newblock {I}nversion of weighted {R}adon transforms via finite {F}ourier
  series weight approximations.
\newblock {\em Inverse Problems in Science and Engineering}, 22(5):787--802,
  2014.

\bibitem[Kun92]{kunyansky1992generalized}
Kunyansky, L.,
\newblock Generalized and attenuated {R}adon transforms: restorative approach
  to the numerical inversion.
\newblock {\em Inverse Problems}, 8(5):809, 1992.

\bibitem[LB73]{lavrientiev1973first}
Lavrent'ev, M. M., Bukhgeim, A. L.,
\newblock {A class of operator equations of the first kind.}
\newblock{ \em Functional Analysis and Its Applications}, 7(4):290-298, 1973.


\bibitem[MQ85]{markoe1985elementary}
Markoe, A., Quinto, E.T.,
\newblock An elementary proof of local invertibility for generalized and
  attenuated Radon transforms.
\newblock {\em SIAM Journal on Mathematical Analysis}, 16(5):1114--1119, 1985.

\bibitem[Natt01]{natterer2001mathematics}
Natterer, F. 
\newblock{The Mathematics of Computerized Tomography}.
\newblock SIAM, 2001.


\bibitem[Nov11]{novikov2011weighted}
Novikov, R.G.,
\newblock {W}eighted {R}adon transforms for which {C}hang's approximate
  inversion formula is exact.
\newblock {\em Russian Mathematical Surveys}, 66(2):442--443, 2011.


\bibitem[Nov14]{novikov2014weighted}
Novikov, R.G.,
\newblock {W}eighted {R}adon transforms and first order differential systems on
  the plane.
\newblock {\em Moscow Mathematical Journal}, 14(4):807--823, 2014.

\bibitem[Qui83]{quinto1983invertibility}
Quinto, E.T.,
\newblock The invertibility of rotation invariant Radon transforms.
\newblock {\em Journal of Mathematical Analysis and Applications},
  91(2):510--522, 1983.
  
\bibitem[Qui83Err]{quinto1983invertibilityerrat}
Quinto, E.T.,
\newblock {The invertibility of rotation invariant Radon transforms. Erratum.}
\newblock{\em Journal of Mathematical Analysis and Applications},
94(2):602-603, 1983.

\end{thebibliography}
\bibliographystyle{alpha}

\end{document}